\documentclass[pra,twocolumn,showpacs,preprintnumbers,amsmath,amssymb]{revtex4}

\usepackage{mathrsfs}
\usepackage{graphicx}
\usepackage{dcolumn}
\usepackage{bm}

\begin{document}

\title{Weak Value and Correlation Function}
\author{Takahiro Sagawa}
\affiliation{Department of Physics, University of Tokyo,
7-3-1, Hongo, Bunkyo-ku, Tokyo, 113-8654, Japan}
\date{\today}

\begin{abstract}
We show that there exists, in quantum theory, a close relationship between the weak value and the correlation function, which sheds new lights on the concept of the weak value. 
\end{abstract}

\pacs{03.65.Ta,03.67.-a}

\maketitle

In this report, we prove an equality which connects two fundamental concepts in quantum theory: the weak value and the correlation function.   The weak value has been a topic of active researches~\cite{Aharonov,Ritchie,Wiseman,Solli,Hall,Wiseman2}, and is related to the foundation of quantum mechanics~\cite{Vaidman}.   
On the other hand, the (symmetric) correlation function is defined as 
${\rm Re} \langle \psi | \hat B (t_2) \hat A (t_1) | \psi \rangle$, where ``${\rm Re}$" means ``the real part", $| \psi \rangle$ is a state vector, and $\hat A (t_1)$  and $\hat B (t_2)$ are observables at time  $t_1$ and $t_2$ in the Heisenberg picture.  The correlation function is a useful quantity to characterize the  dynamics of quantum systems, and has played an important role in quantum statistical mechanics such as the linear response theory~\cite{Kubo}. For example, the correlation function is related to the susceptibility by the fluctuation-dissipation theorem.

To introduce the weak value, we consider two observables $\hat A$ and $\hat B$ of a quantum system.  We denote the spectrum decompositions of them as $\hat A = \sum_a a | a \rangle \langle a |$ and $\hat B = \sum_b b| b \rangle \langle b |$, where $a$'s ($b$'s) are the eigenvalues of $\hat A$ ($\hat B$), and $| a \rangle$'s ($| b \rangle$'s) are the corresponding eigenvectors.  Let $\hat P^A_a \equiv | a \rangle \langle a |$ and $\hat P^B_b \equiv | b \rangle \langle b |$ be the projection operators.  We consider the unitary evolution of the system from time $0$ to $t_2$, and suppose that the projection measurement of $\hat B$ is performed  at time $t_2$, and the outcome $b$ is obtained.  Let $| \psi \rangle$ be the initial state, $t_1$ ($0 < t_1 < t_2$) be an intermediate time, and $\hat U_1$ ($\hat U_2$) be the unitary evolution from $0$ to $t_1$ ($t_1$ to $t_2$).  Then the definition of the (real) weak value of $\hat A$ at time $t_1$ with the post-selection $| b \rangle$ is given by
\begin{equation}
{}_{b}\langle A \rangle_{\rm w} \equiv {\rm Re} \frac{\langle b | \hat U_2 \hat A \hat U_1 | \psi \rangle}{\langle b | \hat U_2 \hat U_1 | \psi \rangle}.
\end{equation}

To relate the weak value to the correlation function, we first define a quasi-probability distribution based on the correlation function:
\begin{equation}
{\rm Pr}_\psi (b, a ) \equiv {\rm Re} \langle \psi | \hat P_b^B(t_2) \hat P_a^A(t_1) | \psi \rangle,
\label{probability}
\end{equation}
where $\hat P_b^B(t_2) \equiv \hat U_1^\dagger \hat U_2^\dagger \hat P^B_b \hat U_2 \hat U_1$ and $\hat P_a^A(t_1) \equiv \hat U_1^\dagger  \hat P^A_a  \hat U_1$ are described in the Heisenberg picture.
We stress that this quasi-probability distribution does not correspond to the probability distribution which we can obtain by performing the successive projection measurements of $\hat A$ at $t_1$ and $\hat B$ at $t_2$.  
The quasi-probability ${\rm Pr}_\psi (b, a)$  satisfies that   $\sum_{a} {\rm Pr}_\psi (b, a) =  \langle \psi | \hat P_b^B(t_2) | \psi \rangle \equiv {\rm Pr}_\psi (b)$,   $\sum_{b} {\rm Pr}_\psi (b, a) =  \langle \psi | \hat P_a^A(t_1) | \psi \rangle \equiv {\rm Pr}_\psi (a)$, and $\sum_{a,b} {\rm Pr}_\psi (b, a ) = 1 $. We note that, in the special case that $\hat A (t_1)$ and $\hat B(t_2)$ are commuting,  the quasi-probability ${\rm Pr}_\psi (b, a )$ reduces to the true probability.

We then calculate the quasi-probability under the condition of $b$ as ${\rm Pr}_\psi (a | b) \equiv {\rm Pr}_\psi (b, a )/ {\rm Pr}_\psi (b) = {\rm Re} \langle \psi | \hat U_1^\dagger \hat U_2^\dagger | b \rangle \langle  b | \hat U_2 \hat P^A_a \hat U_1 | \psi \rangle / \langle \psi |  \hat U_1^\dagger \hat U_2^\dagger | b \rangle \langle  b | \hat U_2  \hat U_1 | \psi \rangle$,
and  obtain
\begin{equation}
{\rm Pr}_\psi (a | b) = {}_b\langle P^A_a \rangle_{\rm w}. 
\end{equation}
Averaging the eigenvalue $a$ over the conditional quasi-probability, we have
\begin{equation}
{}_b\langle A \rangle_{\rm w} = \sum_a a {\rm Pr}_\psi (a | b), 
\label{main}
\end{equation}
which is the main result of this report.  The left-hand side of Eq.~(\ref{main}) is the weak value, and the right-hand side is related to the correlation function via Eq.~(\ref{probability}).


In conclusion, we have derived an equality which gives us a new interpretation of the weak value ${}_b\langle A \rangle_{\rm w}$: it is the average of $\hat A$ under the condition of $b$ over the quasi-probability which is  defined via the correlation function.  While we have only considered the real weak value and the symmetric correlation function, we can straightforwardly generalize our result to the complex weak value and the complex correlation function by removing the notation  ``$\rm Re$".

\begin{acknowledgments}
This work was supported by a Grant-in-Aid for Scientific Research (Grant No.\ 17071005), and by a Global COE program ``Physical Science Frontier'' of MEXT, Japan. TS acknowledges JSPS Research Fellowships for Young Scientists (Grant No. 208038).  TS would like to thank Prof. Masahito Ueda and Mr. Yutaka Shikano  for valuable discussions.
\end{acknowledgments}


\begin{thebibliography}{99}
\bibitem{Aharonov} Y. Aharonov, D. Z. Albert, and L. Vaidman, Phys. Rev. Lett. \textbf{60}, 1351 (1988).
\bibitem{Ritchie} N. W. M. Ritchie, J. G. Story, and Randall G. Hulet, Phys. Rev. Lett. \textbf{66}, 1107  (1991).
\bibitem{Wiseman} H. M. Wiseman,  Phys. Rev. A \textbf{65}, 032111 (2002). 
\bibitem{Solli} D. R. Solli, C. F. McCormick, and R. Y. Chiao, S. Popescu, J. M. Hickmann, Phys. Rev. Lett. \textbf{92}, 043601 (2004).
\bibitem{Hall} Michael J. W. Hall, Phys. Rev. A \textbf{69}, 052113 (2004).
\bibitem{Wiseman2} G. J. Pryde, J. L. O'Brien, A. G. White, T. C. Ralph, and H. M. Wiseman, Phys. Rev. Lett. \textbf{94}, 220405 (2005).
\bibitem{Vaidman} L. Vaidman, Found. Phys. \textbf{26}, 895 (1996).
\bibitem{Kubo} R Kubo, M Toda, and N Hashitsume ``\textit{Statistical Mechanics I\hspace{-.1em}I, Nonequilibrium Statistical Mechanics}''  (Springer, Berlin, 1985).
\end{thebibliography}
\end{document}